\documentclass[runningheads]{svmult}

\usepackage{makeidx}   
\usepackage{graphicx}  
\usepackage{subeqnar}  
\usepackage{multicol}  
\usepackage{physprbb}  
\makeindex             

\begin{document}
\title*{Parameters of PNe: constant density versus density distribution}
\toctitle{Parameters of PNe: constant density versus density distribution}
%
%
\titlerunning{Parameters of PNe: constant density versus density distribution}
%
\author{Michaela Kraus}
%
\authorrunning{Michaela Kraus}
%
%
\institute{Astronomical Institute, Utrecht University, Princetonplein 5, 3584 CC Utrecht, The Netherlands}

\maketitle              

\begin{abstract}
We derive the stellar and circumstellar parameters of the Galactic compact 
planetary nebula Hen 2-90 with two models:
the classical constant density nebula model where the emission (especially of 
the forbiden lines) is formed in a sphere or shell of constant density, and a 
model where the emission is formed in a shell with a $r^{-2}$ density 
distribution. Depending on the density $N_{\rm e,0}$ in the second model,
the resulting range of valid values can deviate from the classical model 
results. This seems to be especially true for elemental abundances as is shown 
exemplary for S$^{+}$. We find that the abundance of S$^{+}$ 
is larger for the models with density distribution than for a constant density.
A careful analysis of data with an appropriate input model is advisable when 
determining stellar and circumstellar parameters of planetary nebulae not only 
in our Galaxy.
\end{abstract}

\section{Introduction}
The parameters of a star and its circumstellar material are difficult to
determine, especially if the star itself is hidden by its circumstellar 
material. In this case, the parameters must be derived indirectly,
for instance from the Balmer and other nebular emission lines.
Especially for planetary nebulae (PNe) there exist some simple 
approximations to derive the effective temperature, distance or 
elemental abundances. But these approximations are normally based on the 
assumption of a constant density spherically symmetric nebula. 
Although the theory of interacting stellar winds that shape the nebulae 
and lead to a density distribution rather than to a constant density, 
which has first been proposed by Kwok et al. (1978), is nowadays accepted
and well established, people still use the
assumption of a constant density nebula to derive
stellar and circumstellar parameters (see e.g. Costa et al. 1993; Ali 1999; 
Phillips 2003; Gruenwald \& Viegas 2000; Villaver et al. 2003, Stanghellini 
et al. 2004).
We investigate differences in the stellar and circumstellar parameters
when applying a constant density model and a model with a density distribution.

\begin{table}
\caption{Line integrated fluxes of some optical emission lines from the
Galactic compact planetary nebula Hen 2-90 and continuum flux at the wavelength 
of the H$\beta$ line. The data have been extinction corrected with $A_{\rm V}
= 4.2^{\rm mag}$ as derived from the Balmer line ratio for $T_{\rm e} = 
10^{4}$\,K. The distance has been set to 2\,kpc.}
\begin{center}
\renewcommand{\arraystretch}{1.4}
\setlength\tabcolsep{5pt}
\begin{tabular}{ccl}
\hline\noalign{\smallskip}
Line & $\lambda$ [\AA] & line integrated flux \\
\noalign{\smallskip}
\hline
\noalign{\smallskip}
H$\alpha$        & 6563 & $5.01\times 10^{-10}$ erg s$^{-1}$ cm$^{-2}$ \\
H$\beta$         & 4861 & $1.73\times 10^{-10}$ erg s$^{-1}$ cm$^{-2}$ \\
$[$S{\sc ii}$]$  & 6731 & $1.47\times 10^{-11}$ erg s$^{-1}$ cm$^{-2}$ \\
$[$S{\sc ii}$]$  & 6716 & $6.54\times 10^{-12}$ erg s$^{-1}$ cm$^{-2}$ \\
\noalign{\smallskip}
\hline
\noalign{\smallskip}
continuum        & 4861 & $1.89\times 10^{-24}$ erg s$^{-1}$ cm$^{-2}$ Hz$^{-1}$ \\
\hline
\end{tabular}
\end{center}
\label{line_param}
\end{table}
                                                                                                      
\section{Observations and models}
                                                                          
To determine the stellar and nebular parameters for the Galactic compact 
planetary nebula Hen 2-90 we make use of a set of optical observations 
specified in Table\,\ref{line_param}. These data have been corrected for an 
interstellar extinction of $A_{\rm V} = 4.2^{\rm mag}$ derived from the Balmer line ratio 
for $T_{\rm e} = 10^{4}$\,K.
We concentrate on two specific models: the constant density nebula model
(CDNM) and a density distribution model (DDM). Both models have the following
assumptions in common:
\begin{itemize}
\item spherically symmetric fully ionized shell or nebula,
\item constant electron temperature fixed at $10^{4}$\,K,
\item all lines in Table\,\ref{line_param} are optically thin, and
\item hydrogen follows case B recombination. 
\end{itemize}
In addition, for CDNM we use a constant electron density throughout the 
nebula while for DDM we assume a density distribution of the form
\begin{equation}\label{eldens}
N_{\rm e}(r) = N_{\rm e,0}\frac{R_{0}^{2}}{r^{2}} 
\end{equation}
$N_{\rm e,0}$ is the electron density at the inner edge, $R_0$, of the emission 
shell and is kept as a free parameter. Such a density distribution has been 
found from hydrodynamic models to be valid over large regions within the wind 
interaction zone (see e.g. Steffen \& Sch\"{o}nberner, 2003).

\section{Results}

\subsection{The effective temperature}

\begin{figure}[t]
\begin{center}
\includegraphics[width=1.0\textwidth]{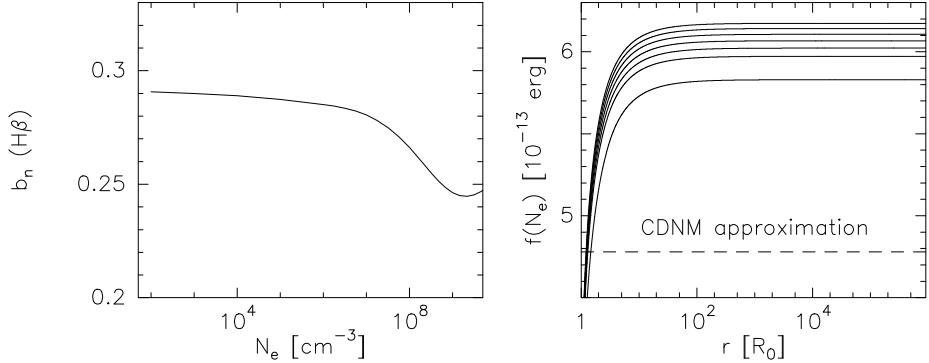}
\end{center}
\caption[]{Left: $b_{\rm n}$ factor for H$\beta$ as a function of 
electron density.
For low values of $N_{\rm e}$ they are fairly constant. 
Right: The function $f(N_{\rm e})$
of Eq.\,(\ref{Zanstratot}) for values of $N_{e,0}$ ranging from 
$10^{2}$\,cm$^{-3}$ (top) to $10^{8}$\,cm$^{-3}$ (bottom). Also
shown is the CDNM value (dashed line)}
\label{bn}
\end{figure}

The effective temperature of the central star of a PN is usually derived 
using the Zanstra method for hydrogen (see e.g. Osterbrock 1989; Pottasch 1984) 
which results in 
\begin{equation}\label{Zanstratot}
\frac{L_{\nu}}{\int\limits_{\nu_{0}}^{\infty}\frac{L_{\nu}}{h\nu}\,d\nu} =
\frac{L_{\nu}}{\frac{L_{{\rm H}\beta}}{h\nu_{{\rm H}\beta}}}\frac{\frac{L_{{\rm H}\beta}}
{h\nu_{{\rm H}\beta}}}{\int\limits_{\nu_{0}}^{\infty}\frac{L_{\nu}}{h\nu}\,d\nu}
= h\nu_{{\rm H}\beta} \frac{F_{\nu}^{\rm obs}}{F_{{\rm H}\beta}^{\rm
obs}}\,f(N_{\rm e})
\end{equation}
The left hand side is the continuum luminosity (e.g. at the wavelength of 
H$\beta$) over the total number of ionizing photons. This is a pure
function of the stellar continuum flux, i.e. a function of effective 
temperature and surface gravity. The right hand side contains the 
(reddening independent) fraction
of observed fluxes times a function $f(N_{\rm e})$ which
is a constant in case of CDNM but a function of electron density
distribution in case of DDM because the $b_{\rm n}$ values that 
describe the deviation from LTE of the Balmer line level populations  
are for high electron densities no longer constant (left panel of 
Fig.\,\ref{bn}). The right panel of Fig.\,\ref{bn} shows  
$f(N_{\rm e})$ for different values of $N_{\rm e}(R_{0})$ ranging from 
$10^{2}$\,cm$^{-3}$ (top) to $10^{8}$\,cm$^{-3}$ (bottom) as a function 
of $r$.
\begin{figure}[b]
\begin{center}
\includegraphics[width=0.7\textwidth]{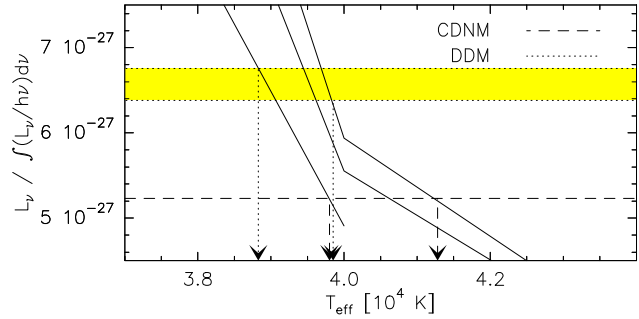}
\end{center}
\caption[]{Range of effective temperatures derived with the Zanstra 
method for H in case of CDNM (dashed arows) and DDM (dotted arrows) for the 
$N_{\rm e,0}$ values of Fig.\,1. The top boundary of the shaded region is 
for $N_{\rm e,0} = 10^{2}$\,cm$^{-3}$, the bottom boundary for $N_{\rm e,0} 
= 10^{8}$\,cm$^{-3}$. The solid lines are Kurucz models for the stellar 
$\log g = 5.0; 4.5; 4.0$ (from right to left)} 
\label{Zanstra}
\end{figure}
The range of possible effective temperature and surface gravity
combinations of the central star derived with the Zanstra method for H is shown in Fig.\,\ref{Zanstra} where we
used Kurucz model atmospheres (Kurucz 1979) to calculate the left
hand side of Eq.\,(\ref{Zanstratot}). While CDNM gives values in the 
range of $39\,800\le T_{\rm eff}/{\rm K} \le 41\,300$, the range found 
for DDM is shifted to lower values covering the range 
$38\,800\le T_{\rm eff}/{\rm K} \le 39\,850$; the lowest temperature 
is for the lowest (choosen) $N_{\rm e,0}$ value of $10^{2}$\,cm$^{-3}$.
The temperatures found with the two models are not extremely different; 
both methods result in a spectral type O for the central star of this
compact planetary nebula.

\begin{figure}[b]
\begin{center}
\includegraphics[width=1.0\textwidth]{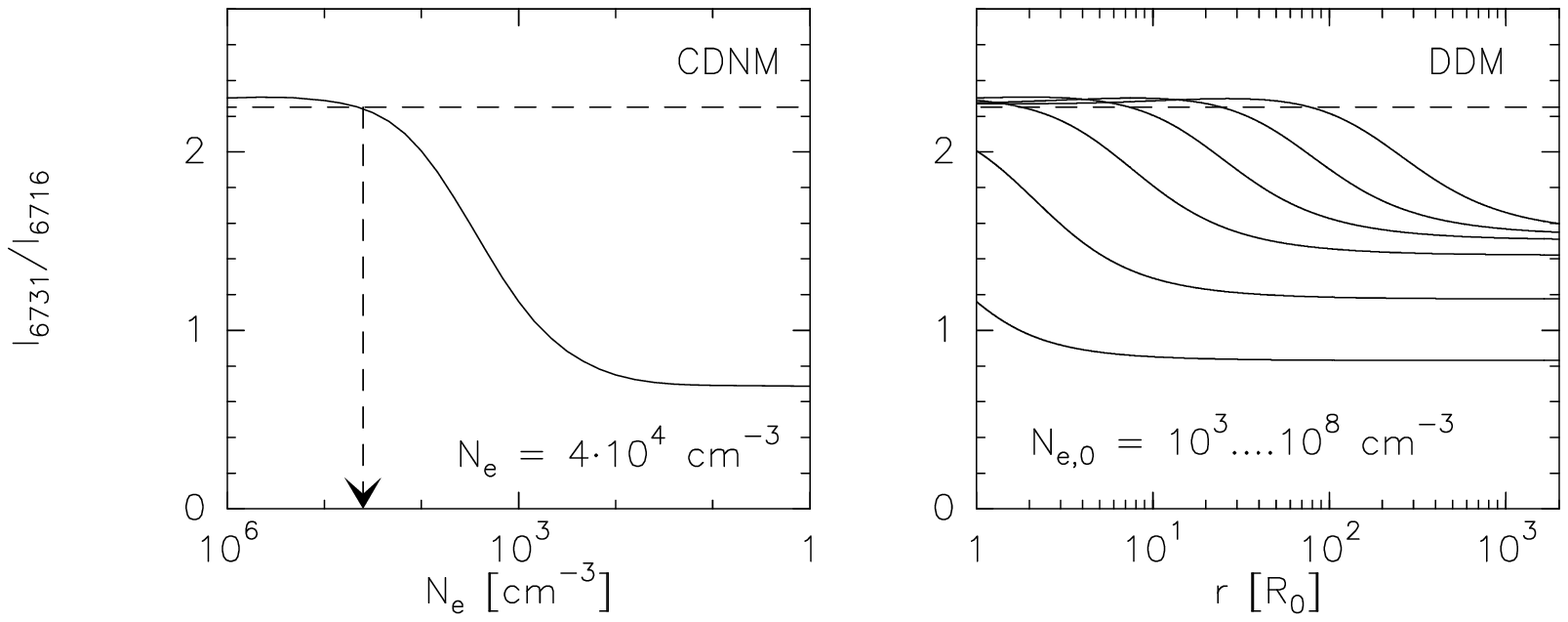}
\end{center}
\caption[]{Left: Constant electron density in CDNM derived from the observed
(dashed line) sulfur line ratio. Right: The sulfur line ratios in case of DDM 
for $N_{\rm e,0}$ ranging from $10^{3}$\,cm$^{-3}$ (lower left)
to $10^{8}$\,cm$^{-3}$ (upper right). The observed ratio (dashed line) limits 
$N_{\rm e,0}$ to values 
$\mathrel{\hbox{\rlap{\hbox{\lower4pt\hbox{$\sim$}}}\hbox{$>$}}}
10^{5}$\,cm$^{-3}$. The crossing point between model and observation defines 
the outer edge of the emitting shell for each model}
\label{lineratio}
\end{figure}

\subsection{Elemental abundances}

\begin{figure}
\begin{center}
\includegraphics[width=1.0\textwidth]{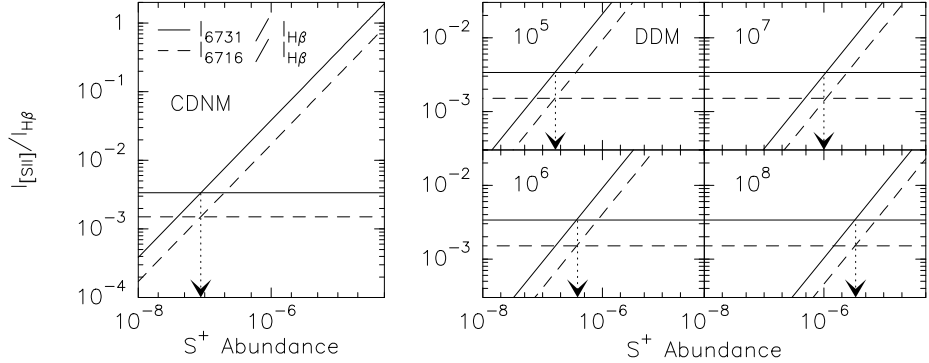}
\end{center}
\caption[]{S{\sc ii} over H$\beta$ intensity ratio as a function of 
S$^{+}$ abundance for CDNM (left) and DDM (right) for the different possible
$N_{\rm e,0}$ values as indicated. Horizontal lines indicate
observed ratios, arrows indicate the derived S$^{+}$ abundance. All 
abundances found with DDM are clearly larger than that for CDNM}
\label{abundance}
\end{figure}

We take for example the [S{\sc ii}] lines 
and calculate for both models the intensity ratio of a [S{\sc ii}] line
over H$\beta$ (for details see e.g. Pottasch et al. 2003). For CDNM we need to know the  
constant electron density. This can be found from the sulfur line 
ratio (left panel of Fig.\,\ref{lineratio}) and turns out to be 
$4\times 10^{4}$\,cm$^{-3}$. The resulting ratio of the two S{\sc ii}
lines over H$\beta$ is shown in the left panel of Fig.\,\ref{abundance}.
From this plot, a S$^{+}$ abundance of $8.75\times 10^{-8}$ is found.

In case of DDM the observed
sulfur line ratio defines the lower limit of $N_{\rm e,0}$
(right panel of Fig.\,\ref{lineratio}) which is about $10^{5}$\,cm$^{-3}$.
Models with $N_{\rm e,0} < 10^{5}$\,cm$^{-3}$ cannot reproduce the observed 
line ratio of 2.25.
In addition, the crossing point of the theoretical line ratio curve with the 
observed ratio defines the outer edge (in terms of $R_0$) of the emission shell
for each density distribution model. This outer edge has to be taken into 
account when caclulating the intensity ratio of S{\sc II} over H$\beta$. The
results are shown in the right panel of
Fig.\,\ref{abundance}. We find that the S$^{+}$ abundance values calculated
with DDM are all larger than the value found with CDNM, with the lowest value of
$1.64\times 10^{-7}$ for $N_{\rm e,0} = 10^{5}$\,cm$^{-3}$ which is about twice the CDNM value. 

\section{Conclusions}

We derived the effective temperature and the S$^{+}$ abundance for the 
Galactic compact PN Hen 2-90 using the constant density nebula approach and a 
shell model with a $r^{-2}$ density distribution. We find that the effective 
temperature of the central star derived with DDM is slightly (but not significantly) lower than with CDNM, classifying the star as spectral type O. 
The most interesting result is, however, that the S$^{+}$ 
abundance found with DDM is definitely larger than the value found with CDNM. 
Even though we neglected additional ionization stages, this result might hold
for all abundances derived from forbidden lines and might solve the problem of 
the underabundances sometimes found in PNe.

%

%


\begin{thebibliography}{12.}
\addcontentsline{toc}{section}{References}

\bibitem{Ali}
        A. Ali: New Astronomy \textbf{4}, 95 (1999)
\bibitem{Costa}
        R.D.D. Costa, J.A. de Freitas Pacheco, W.J. Maciel: A\&A 
        \textbf{276}, 184 (1993)
\bibitem{Gruenwald}
        R. Gruenwald, S.M. Viegas: ApJ \textbf{543}, 889 (2000)
\bibitem{Kurucz}
        R.L. Kurucz: ApJS \textbf{40}, 1 (1979)
\bibitem{Kwok}
        S. Kwok, S., C.R. Purton, P.M. FitzGerald: ApJ \textbf{219}, 
        L\,125 (1978)
\bibitem{Osterbrock}
        D.E. Osterbrock: \emph{Physics of gaseous nebulae and active
        galactic nuclei} (University science books, Mill Valley 1989)
\bibitem{Phillips}
        J.P. Phillips: MNRAS \textbf{344}, 501 (2003)
\bibitem{Pottasch84}
        S.R. Pottasch: \emph{Planetary nebulae} (D.Reidel Publishing 
        Company, Dordrecht 1984)
\bibitem{Pottasch03}
        S.R. Pottasch at al: A\&A \textbf{409}, 599 (2003)
\bibitem{Stanghellini}
        L. Stanghellini et al: ApJ \emph{in press}, astro-ph/0411631 (2004)
\bibitem{Steffen}
        M. Steffen, D. Sch\"{o}nberner: \emph{Planetary Nebulae: Their 
        Evolution and Role in the Universe}, IAU Symposium 209, ed: 
        S. Kwok, M. Dopita, R. Sutherland, (ASP, San Francisco, 2003), 439
\bibitem{Villaver}
        E. Villaver, L. Stanghellini, R.A. Shaw: ApJ \textbf{597}, 298 
        (2003)


\end{thebibliography}
\end{document}